\documentclass[12pt]{article}
\usepackage{amsfonts,latexsym,fullpage,amsmath,latexsym,amssymb,epsf}

\date{June 03, 2003}

\newcommand{\cW}{{\cal W}}

\newcommand{\beq}{\begin{equation}}
\newcommand{\eeq}{\end{equation}}
\newcommand{\beqy}{\begin{eqnarray}}
\newcommand{\eeqy}{\end{eqnarray}}

\newtheorem{Definition}{Definition}
\newtheorem{Lemma}{Lemma}
\newtheorem{Theorem}{Theorem}

\newenvironment{Proof}{{\it Proof: \,}}{$\Box$ \vspace{0.3cm}}

\newenvironment{Definition*}{{\bf Definition}}{}

\def\C{{\mathbb{C}}}
\def\Z{{\mathbb{Z}}}
\def\R{{\mathbb{R}}}

\newcommand{\cH}{{\cal H}}

\begin{document}

\title{Synchronizing quantum clocks with classical one-way communication:
Bounds on the generated entropy}

\author{Dominik
Janzing\thanks{e-mail: 
janzing@ira.uka.de} \,\, and Thomas
Beth \\ \small Institut f{\"u}r Algorithmen und Kognitive Systeme,
Universit{\"a}t Karlsruhe,\\[-1ex] \small Am Fasanengarten 5,
D-76\,131 Karlsruhe, Germany}

\maketitle

\abstract{We describe separable joint states on bipartite
quantum systems 
that cannot be prepared by any thermodynamically
reversible classical one-way communication protocol.
We argue that
the  joint state of two synchronized microscopic
clocks is always of this type when it is considered
from the point of view of an  ``ignorant''
observer who is not synchronized with the other two
parties. 

We show that the entropy generation of a classical
one-way synchronization protocol is at least
$\Delta S = \hbar^2/(4\Delta E \Delta t)^2$ 
if $\Delta t$ is the time accuracy of the synchronism
and $\Delta E$ is the energy bandwidth of the clocks.
This dissipation can only be avoided if  the
common time of the microscopic clocks 
is stored by an additional  classical clock.

Furthermore, we give a similar bound on the entropy cost
for  {\it resetting} synchronized clocks by a classical one-way protocol.
The proof relies on observations of Zurek on the thermodynamic
relevance of {\it quantum discord}.
We leave it as an open question whether classical multi-step  protocols
may perform better.

We discuss to what extent our results imply problems
for classical  
concepts of reversible
computation when the energy of timing signals is close
to the Heisenberg limit.}

\section{The thermodynamic advantage \\ of quantum information transfer}

\label{Adv}

Among the most important properties of quantum channels is their
ability to create entangled states in a bipartite or
multipartite  system. It is 
well-known that every non-entangled (i.e. separable) joint state
can be prepared using local quantum operations on each subsystem
and classical communication among them.
However, although it is possible to prepare all separable joint states
with classical communication it may nevertheless be advantageous
to use quantum communication for thermodynamic reasons.
Bennett et al. \cite{BennettNonlocal} considered the following problem:
Alice and Bob receive each a classical message $i$. 
 Alice is instructed to
prepare the quantum state  $|\alpha_i\rangle$ 
when she receives the message $i$ and Bob should prepare the state
$|\beta_i\rangle$ whenever he receives the message $i$.
They found a set of tensor product 
states $|\psi_i\rangle:=|\alpha_i\rangle \otimes |\beta_i\rangle$ with
the property that all $|\psi_i\rangle$ 
are mutually orthogonal but neither all
states $|\alpha_i \rangle$ nor all states $|\beta_i\rangle$ 
are orthogonal.  The message $i$ 
received by Alice and Bob  is
certainly represented by any physical system and can hence be modeled
without loss of generality by mutually orthogonal quantum states 
$|i\rangle$.
Hence Alice and Bob share the state $|i\rangle \otimes |i\rangle$ after
they have received the instruction. Clearly there is a unitary transformation
$U$ acting on the composed Hilbert space transforming $|i\rangle \otimes 
|i\rangle$ into $|\alpha_i\rangle \otimes |\beta_i\rangle$. But there 
are no {\it local} unitary transformations 
$U_A$ and $U_B$  (independend of $i$)  for Alice and 
Bob,  converting $|i\rangle$ into $|\alpha_i\rangle$
and $|\beta_i\rangle$. 
The authors of \cite{BennettNonlocal} 
conjecture that Alice and Bob necessarily have to use thermodynamical
irreversible operations in order to achieve their tasks.
Note that, in their setting, the task is not to prepare a density matrix
of the form
\[
\sum_i p_i |\alpha_i \rangle \langle \alpha_i | 
\otimes |\beta_i \rangle \langle \beta_i |\,,
\]
it is rather to prepare the state $|\alpha_i\rangle \otimes |\beta_i\rangle$
whenever the message was $i$. This problem does only make
sense when a third party keeps the message $i$ in mind, i.e., if
one has prepared the tripartite density matrix
\[
\sum_i p_i |i\rangle \langle i|\otimes 
|\alpha_i \rangle \langle \alpha_i | 
\otimes |\beta_i \rangle \langle \beta_i | \,,
\]
where the left component is a memory for the message. From this point of view,
they consider the preparation of {\it tripartite} density matrices.

In this paper we consider, in contrast to the setting above,
the preparation of {\it bipartite} density matrices.
The type of density matrices considered here appears naturally
when two synchronized ``microscopic clocks'' (quantum dynamical systems)
are considered from the point of view of an observer without
clock or without knowledge of the common time of the other two parties.
These bipartite states have the property that they have no
decomposition into locally distinguishable product states.
States of this type have  already been considered by Ollivier and 
Zurek \cite{ZurekDiscord}. The authors observed that 
3 classically equivalent ways to define mutual information 
differ in the quantum case.
They called the  difference  {\it quantum discord}. 
Zurek observed \cite{ZurekDemon} the thermodynamic relevance of this quantity:
He considered  how much entropy has to be transferred 
into the environment when both parties 
want to obtain a pure state provided they are restricted to
{\it classical} one-way communication from Alice to Bob.
He showed that the thermodynamic cost is lower when they are allowed to use 
quantum communication and the difference of the entropy generation
is exactly  the quantum discord.
In Sections~\ref{Obvious} and \ref{Bound} 
 we consider in some sense the inverse problem to {\it prepare}
correlations of this kind with classical one-way communication
and given a bound on the entropy generation.
In the following we will define the class of states that we consider
and explain why they should be considered as {\it synchronized clocks}.
The formal setting for defining synchronization 
is mostly taken from our paper
\cite{clock}.
In Section~\ref{Discord} we consider the problem to {\it resolve} the
correlations of the joint states (``to reset'' the synchronized clocks)
using classical one-way communication and prove a lower bound on the
 quantum discord of the two-clocks state.
In Section \ref{Explain} we  explain why our model assumptions
seem to appear naturally in real applications.
In Section \ref{Imply} we discuss to what extent our results
imply serious constraints or problems for classical concepts of
extremely low power computation as soon as the signal energy 
of clock signals is close to the Heisenberg limit.

\section{How to define synchronism \\ of microscopic clocks}

\label{SynIntro}

In the first place we note that synchronization involves
clocks in their broadest sense, namely physical systems that are 
evolving in time. Consider for instance a classical system which has the
unit circle $\Gamma$ 
in $\R^2=\C$ as its ``phase space'' and the dynamics
is just the rotation 
\[
z \mapsto z \exp(i\omega t)\,.
\]
This would be a very primitive clock since it shows the time only
up to multiples of the period $T=2\pi/\omega$. However, at least 
this is done perfectly.

Now consider two parties, $A$ and $B$ (Alice and Bob) each having
a clock of this kind. Then we may say  that both are synchronized with
the external time. 

We want to define synchronization between Alice and Bob in a way that
does not refer to an absolute (external) time but 
formalizes only the fact that both clocks agree perfectly.
An observer who does not know the time $t$ will not
realize
that both clocks are in the position
$z=\exp(i\omega t)$ at time $t$, he will only observe that the positions
$z_A$ and $z_B$ 
of both ``pointers'' agree.
In the language of probability theory, he may describe this fact by
a measure $\delta $ on $\Gamma\times \Gamma$ 
which is defined by
\[
\delta (M \times K) = \lambda (M\cap K)
\]
where $\lambda$ is the normalized Lebesgue measure on the 
unit circle. If $z_A$ and  $z_B$ 
agree always up to a well-defined phase difference 
$\phi$ the clocks are also synchronized.
Formally, we define:

\begin{Definition}
Two classical clocks (with equal frequency $\omega$) 
which are described by rotating pointer on the 
unit circle are called {\it perfectly synchronized} 
if the external observer 
describes the expected pointer positions  $z_A$ and $z_B$ of their 
clocks by a joint probability distribution of the form
\[
\delta_\phi
\]
with
\[
\delta_\phi(M\times K):=\lambda (M\cap (K\exp(i\phi)))
\]
and 
$\phi$  is the phase difference which is known
to the external observer.
\end{Definition}

An imperfect synchronism may, for instance, be a joint measure
described by a probability density
\[
p(z_A,z_B)=f(z_Az^{-1}_B)
\]
where $f$ is an appropriate continuous function with support essentially at
the point $1$ or any other point $\exp(i\phi)$ where $\phi$ is the
expected phase difference between the clocks.
These remarks should only show why synchronized classical
clocks may be described by a time invariant joint measure with
non-trivial correlations.

An interesting kind of synchronism is given if 
both parties have ``quantum clocks'', i.e., quantum systems
with non-trivial time evolution. Consider for instance the case
that each one has a two level system where $|0\rangle$ and $|1\rangle$
denote the lower and upper level, respectively. Each is evolving in time
according to 
\[
|\psi_t\rangle:=\frac{1}{\sqrt{2}}(|0\rangle + \exp(-i\omega t)|1\rangle)\,.
\]
From the point of view of the ``ignorant'' observer
the bipartite system is in the joint state
\[
\sigma:=\frac{1}{T}\int_0^T\rho_t \otimes \rho_t \,dt \,\,\,
\hbox{ with } \,\,\, \rho_t:=|\psi_t \rangle \langle \psi_t|\,.
\]
In some sense, this is a perfectly synchronized system since
the phases of both two-level systems agree exactly.
On the other hand, none of both can read out the phase of its system
and so they cannot use this synchronism as resource to obtain
perfectly synchronized classical clocks. But they could obtain imperfectly
synchronized classical clocks by performing measurements on their systems
and adjust   their clocks according to the measurement results.
The most natural way to do this is to apply a covariant 
positive operator
valued measure $(M_t)_{t\in [0,2\pi/\omega]}$ satisfying the 
covariance 
condition 
\[
\exp(iHt)M_s \exp(-iHt) =M_{s+t}\,,
\]
where $H:=diag (1,0)$ is the Hamiltonian of the two-level system.
The probability density
for obtaining 
the time $t'$ if the true time is  $t$ is now given by
\[
p(t'|t):=tr(\rho_t M_{t'})\,.
\]
The observer who only notices these two classical clocks which
have been adjusted according to the estimated times will only
note that the probability that $A$ has the time $t_A$ and $B$ has the time
$t_B$ is given by
\[
q(t_A,t_B)=\int_0^T tr((\rho_{t} \otimes \rho_{t}) 
M_{t_A} \otimes M_{t_B})
dt
\]
If the POVM $(M_t)$ is not completely useless, i.e., 
if it is chosen such that
$tr(\rho_t M_{t'})$ is not constant for all $t,t'$ 
the probability distribution  $q$ contains some synchronism in the
following
sense: 
There are some correlations between the time estimations of 
of $A$ and $B$. 
A large number of copies of such pairs
of ``weakly synchronized'' clocks are as worthy as one well synchronized
pair. Considerations of this kind are made precise in
\cite{clock} where we have defined as so-called ``quasi-order of clocks''
and a ``quasi-order of synchronism''.
To rephrase the relevant part of this concept
we have to define a clock:

\begin{Definition}
A (quantum or classical) 
clock is a  physical system with 
non-trivial dynamics. 
We denote it as a pair
$(\rho,\alpha)$ where $\rho$ is the state at the time $0$ and
$\alpha:=(\alpha_t)_{t\in \R}$ 
denotes the time evolution, i.e., $\alpha_t(\rho)$ is the
state at the time $t$.
In the classical case $\rho$ is a probability distribution on the
phase space $\Omega$ 
 of the system and $\alpha_t: \Omega \rightarrow \Omega$ 
is a flow in the phase space 
shifting this measure.

In the quantum case $\rho$ is a density matrix  and 
$\alpha_t(\rho):=\exp(-iHt)\rho \exp(iHt)$ is the  
time evolution according to the Hamiltonian $H$.
\end{Definition}

The ability of the clock $(\rho,\alpha)$ to show the time
is given by the distinguishability of the states
\[
\rho_t:=\alpha_t (\rho)\,.
\]

In \cite{clock} we have developed a formal setting and theory to
classify clocks with respect to their quality.
A unifying framework describing quantum and 
classical physical systems is given by the $C^*$-algebraic approach 
\cite{clock}.

To define synchronism of clocks formally we state that two physical systems
with non-trivial separate time evolutions $\alpha$ and $\beta$ 
 are to some extent synchronized
if and only if their joint state is correlated in such a way that 
the correlations carry some
information about a common time. The following setting
includes also the trivial case that there is no common timing information:

\begin{Definition}
A synchronism is a triple $(\sigma,\alpha,\beta)$ where
$\sigma$ is the joint state of
a bipartite physical system,  $\alpha$ 
and $\beta$ are the time evolutions corresponding
to the first and second system, respectively
and $\sigma$ is invariant under $\alpha_t \otimes \beta_t$. 
\end{Definition}

Then we formalize whether a synchronized pair of clocks
 can be used as resource
to synchronize another pair of clocks sufficiently:

\begin{Definition}
A bipartite system  $(\sigma,\alpha,\beta)$ 
is at least as good synchronized as
$(\tilde{\sigma},\tilde{\alpha},\tilde{\beta})$ if
there is a completely positive trace preserving map
$G$ satisfying the covariance condition
\[
G \circ (\alpha_t \otimes \beta_{-t}) = (\tilde{\alpha}_t \otimes 
\tilde{\beta}_{-t}) \circ G
\]
such that
\[
G (\sigma)= \tilde{\sigma}
\]
We say $(\sigma,\alpha,\beta)$ is sufficient as resource 
to prepare $(\tilde{\sigma},\tilde{\alpha},\tilde{\beta})$
\end{Definition}

Note that the covariance condition ensures that the conversion process
does not refer to additional synchronized clocks \cite{clock}.
The least elements in this quasi-order are given by those synchronisms
$(\sigma,\alpha,\beta)$   which satisfy
\[
(\alpha_t \otimes \beta_{-t})(\sigma)= \sigma\,.
\]
This is intuitively plausible since it does not require
any synchronization procedure or synchronized clocks to prepare
a system that is invariant with respect to relative time translations among
both parties.
Formally it is also easy to see since one can 
define $G$ such that it maps every state of the considered
resource system
onto the state $\sigma$. 
This satisfies clearly the covariance condition independent of 
the time evolution of the resource system.

Consequently, we define:

\begin{Definition}
A bipartite system 
 $(\sigma,\alpha,\beta)$ is not synchronized
if $(\alpha_t \otimes \beta_{-t}) (\sigma)=\sigma$. Otherwise
we call it ``to some extent synchronized''.
\end{Definition}

Now we briefly 
consider the question
how the quality of a synchronism can be measured.
Assume Alice and Bob perform measurements on their clocks in order to
obtain information about the time. It is natural to restrict the attention
to time-covariant measurements. Alice's and Bob's 
measurements are described  by positive operator valued measurements
$(M_t)_{t\in [0,T)}$ and $(N_s)_{s\in [0,T)}$, respectively.
They satisfy $M_t=\alpha_{-t}(M_0)$ and
$N_s=\beta_{-s}(N_0)$.
We define the mean quadratic 
deviation between Alice's and Bob's measurement result
by 
\[
D:=\int_0^T\int_0^T (s-t)^2 \,\,tr(\sigma (M_t\otimes N_s))\,\, ds dt\,.
\]
Since $s,t$ are only defined modulo $T$ 
the expression
  $(s-t)^2$ is to be understood as
\[
\min_{l\in \Z} \{ (s-t+ lT)^2\} \,.
\]
In abuse of notation we denote 
\[
\Delta t:=\sqrt{D}\,,
\]
and call $\Delta t$  the {\it standard time deviation}.

Now we restrict our attention to systems which are purely quantum, i.e.,
Alice's and Bob's clocks  are moving according to their Hamiltonians
$H_A$ and $H_B$ acting on the
 Hilbert spaces
$\cH_A$ and $\cH_B$, respectively. 

After the synchronization
Alice and Bob share a joint density matrix on 
$\cH_A \otimes \cH_B$.
The following quantity will be a useful measure for 
the degree of synchronization since
it quantifies
the non-invariance of the joint state with respect
to the relative time translation $\alpha_{t/2}\otimes \beta_{-t/2}$:

We consider
the trace-norm of the derivative of the joint state
with respect to relative time translation, i.e.,
\begin{eqnarray}\label{durchziehen}
\|\frac{d}{dt} (\alpha_{t/2} \otimes \beta_{-t/2}) (\rho)\|_1
&=&\frac{1}{2}\|[H_A\otimes {\bf 1} -{\bf 1} \otimes H_B, \sigma]\|_1
\\&=&
\|[H_A \otimes {\bf 1} , \sigma]\|_1=\| [{\bf 1}\otimes H_B,\sigma]\|_1
\nonumber
\,.
\end{eqnarray}
These equations follow easily from the  invariance
of $\sigma$ with respect to the dynamical evolution which is 
generated by the 
Hamiltonian $H:=H_A \otimes {\bf 1} + {\bf 1}\otimes H_B$. 
The quantity in eq.~(\ref{durchziehen}) 
has a less intuitive meaning than the
standard time deviation but it will help to prove our main theorem
in Section \ref{Bound}.  
The following Lemma draws a connection to the standard time deviation.

\begin{Lemma}\label{IntDelta}
Let the standard deviation 
$\Delta t$ of a synchronism $(\sigma,\alpha,\beta)$ 
be much smaller than the period $T$ of the clocks
(Here we assume
$\Delta t\leq T/12$ for technical reasons to 
 get a simple proof).
Then
we have the following inequality:
\[
\frac{1}{4\Delta t} \leq \|\frac{d}{dt} 
(\alpha_{t/2} \otimes \beta_{-t/2}) (\sigma)\|_1\,.
\]
\end{Lemma}
\begin{Proof}
Define the observable
\[
A:= \int_{|s-t| \leq 2\Delta t} M_t \otimes N_s \,\, ds dt
-\int_{|s-t| \geq 2\Delta t} M_t \otimes N_s \,\, ds dt\,.
\]
Its operator norm is not greater than $1$.
Consider $s-t$ as a random variable with values in $[-T/2,T/2)$.
The generalized Tschebyscheff inequality  
states that for every random variable 
$X$ the event $|X|\geq \epsilon$ occurs at most 
with probability 
$E(X^2)/\epsilon^2$ when $E(X^2)$ denotes the expectation value
of $X^2$.
We conclude  that $|s-t|$ exceeds $2 \Delta t$ 
at most with probability $1/4$ 
(note that here $|s-t|$ is to be understood as
the minimum $|s-t -lT|$ for $l\in \Z$).
This implies 
\[
tr (\sigma A) \geq 3/4 -1/4 =1/2\,.
\]
Now consider the state that is obtained from $\rho$  by 
relative time translation of the amount 
$4\Delta t$. Set $r:= \Delta t$ and
\[
\tilde{\sigma}:=(\alpha_{2r} \otimes\beta_{-2r}) (\sigma) \,.
\]
Due to the covariance of the operators $M_t$ and $N_s$ 
and the condition $\Delta t \leq T/12$ 
we know that
with probability at least $3/4$ the values $s$ and $t$ satisfy
\[
-6r\leq s-t   \leq -2r\,.
\]
This implies obviously 
\[
tr( \tilde{\sigma} A) \leq -3/4 + 1/4= -1/2\,.
\]
Since the expectation value 
\[
tr((\alpha_{t/2} \otimes\beta_{-t/2}) (\sigma) A )
\]
decreases from $1/2$ to $-1/2$ within an interval of length $4r$ 
the average derivative of 
\begin{equation}\label{Der}
\frac{d}{dt} tr(\alpha_{t/2} \otimes \beta_{-t/2} (\sigma) A)
\end{equation}
is less or equal to
$-1/(4r)$ on this interval. Note that the modulus of expression (\ref{Der}) 
is bounded by 
\begin{equation}\label{Der2}
\|\frac{d}{dt} \alpha_{t/2} \otimes\beta_{-t/2} (\sigma) \|_1\,,
\end{equation}
which is constant for all $t$. Hence expression (\ref{Der2}) 
has to be at least
$1/(4r)$.
\end{Proof}

In the following section we will prove a lower bound on the entropy
increase based on Lemma \ref{IntDelta}. We do not claim that the bound
is tight since it uses inequalities connecting trace-norm distances between
quantum states with relative entropies. However, we were not able to
find tighter bounds in this general setting.

\section{A simple classical synchronization protocol}

\label{Obvious}

The following scheme shows a straightforward method to achieve 
synchronization for a simple type of quantum clocks 
on Hilbert spaces of arbitrary finite dimension.
Let the clocks of Alice and Bob each be described 
by the Hilbert space $\C^n$ and the Hamiltonian
be $H:=diag (0,1,\dots,n-1)$.
Let $|\psi\rangle$  be a uniform superposition of basis states $|j\rangle$ 
in $\C^n$:
\begin{equation}\label{Superp}
|\psi\rangle :=\frac{1}{\sqrt{n}}\sum_{j=0}^{n-1} |j\rangle \,.
\end{equation}
Let $\rho_t$ be the density matrix obtained from $|\psi\rangle \langle \psi|$ 
after the time $t$, i.e.,
\[
\rho_t := e^{-iHt}  |\psi\rangle \langle \psi| e^{iHt}\,.
\]
If these clocks are optimally synchronized Alice and Bob share the joint
state 
\[
\sigma:= \int_0^{2\pi} \rho_t \otimes\rho_t \,dt \,.
\]
Due to a theorem of 
Carath\'{e}odory \cite{Gruenbaum} this state can also be obtained by
a finite convex combination of product states.
Elementary  Fourier analysis arguments show that $\sigma$ can also
be obtained by 
\begin{equation}\label{State}
\sigma=\frac{1}{2n-1} \sum_{j=1}^{2n-2} \rho_{t_j} \otimes \rho_{t_j} \,,
\end{equation}
with $t_j:=2\pi j/ (2n-1)$.

We are looking for a protocol with the following properties:

\begin{enumerate}

\item Alice and Bob start with a product state. Both are allowed 
to perform any arbitrary local operations on their physical systems.
Their physical systems may be of  arbitrary dimension
and they have unrestricted 
access to ancilla systems. The only physical systems with non-trivial
dynamics 
are given by one Hamiltonian quantum dynamical system 
on Alice's side (Alice's quantum clock), one Hamiltonian 
quantum dynamical 
system on Bob's side (Bob's quantum clock), and a classical clock
on Alice's side. All the other systems have trivial time evolution.

\item Alice sends Bob a package
consisting of the classical clock
and a memory with some additional information.

\item Bob receives the clock and the memory and keeps both. Then he is 
allowed to implement any transformation on the extended system
consisting of his quantum clock and the received package.

\item At the end of the protocol the joint state of Alice's and Bob's
quantum clocks should be uncorrelated with the classical clock.
Otherwise there would be trivial thermodynamically reversible way
to achieve synchronization using an ideal ``circle clock''
as introduced at the beginning of Section \ref{SynIntro}:
Alice performs the transformation
$
\exp(-iHt)
$
if the classical clock has the time $t$. She sends the classical clock 
to Bob and he implements $\exp(-iHt')$ according to the actual time $t'$.
For the external observer the result is a 
tripartite joint state of the three clocks.
In Section \ref{Explain} we will explain in detail why we exclude protocols
where both quantum clocks  are afterwards committed to a classical clock.
The idea is that we have applications in mind where 
the classical  clock is a signal which is {\it absorbed} 
by the receiver (Bob) and hence does not exist any longer.
\end{enumerate}

The following protocol satisfies all these requirements and
prepares
the state (\ref{State}):
We assume that Alice has a classical memory with $2n-1$ possible states.
We assume that it is not initialized, i.e., it is in the
 mixed state
\[
\gamma:= \frac{1}{2n-1}\sum_{j=0}^{2n-2} |j\rangle \langle j |
\]
Her clock is assumed to be in the state $|\psi \rangle \langle \psi|$
(defined as in eq.~(\ref{Superp})).
Then she performs a unitary transformation conditional on the state
of the memory. If the memory state is $|j\rangle \langle j|$ she  
implements
the unitary operation 
\begin{equation}\label{discreteHam}
\exp(-iH \frac{2\pi \,j}{2n-1})
\end{equation}
on the quantum clock.
Furthermore we assume that she has a classical clock.
If it shows the time $t$ she implements 
\begin{equation}\label{Op1}
\exp(-iHt)
\end{equation}
on her quantum  clock.
Afterwards she sends the memory and the classical clock to Bob.
When he receives both he implements
\begin{equation}\label{Op2}
\exp(-iHt')
\end{equation}
 when the classical clock shows the time $t'$. 
The operations \ref{Op1} and \ref{Op2} ensure that the 
joint state obtained at the end of the protocol
does not depend on  the time the message needs to reach Bob.
Then 
he implements
also the conditional transformation~\ref{discreteHam}  
whenever the message is $j$.
Now the reduced state of the system
consisting of Alice's and Bob's clocks is already the desired state 
$\sigma$ but the two clocks are still correlated with the memory.
So far, the protocol is thermodynamically reversible and the
joint state of Alice's and Bob's quantum clock and the memory 
is uncorrelated with the classical clock as desired.

However, the joint state of memory and both clocks evolves in time.
Hence it is only pure from the point of view of somebody who
knows the time which has passed by since the synchronization
has taken place. From the point of view of
an ignorant observer  the protocol is hence only reversible when
the joint state is still correlated with an additional classical clock
showing the time that has been passed by.
Here we do not allow this and have hence entropy increase by
``forgetting the time''. The reduced state $\gamma$ 
of the memory is obviously
stationary since the memory is a stationary system by assumption.
The reduced state of the composed system consisting of both clocks
is also stationary since it is the desired state $\sigma$.
Forgetting the time destroys the correlations between memory and
both clocks and leads to the state
\[
\sigma\otimes \gamma\,.
\]
Hence the entropy increase is exactly the mutual information
 between memory and both clocks. The entropy of the memory
is $S(\gamma)$, the entropy of the joint state 
of Alice's and Bob's clock is
$S(\sigma)$. 
The entropy of the joint state (before the time has been
forgotten) is $S(\gamma)$. Hence the mutual information is $S(\sigma)$.
It can be calculated as follows. Let $\cH_j$ be the eigenspace of
$H_A \otimes {\bf 1} + {\bf 1}\otimes H_B$ corresponding to the eigenvalue $j$.
They have dimension $j+1$ for $j\leq n$ and dimension $2n-j+1$ for
$j\geq n$. Since $\sigma$ is stationary it is block diagonal with respect
to this decomposition into subspaces. 
By elementary Fourier analysis one can see that all entries of each 
block matrix
are $1/n^2$. The eigenvalues of a matrix of dimension $d$ 
which has only $1$ as entries  are
given by $0,0,\dots,0,d$. Hence the eigenvalues of $\sigma$ are
\[
\frac{j}{n^2}
\]
for $j=0,1,\dots,n$. For each $1\leq j\leq n-1$ the 
corresponding eigenvalue occurs twice, the eigenvalue $1/n$ occurs only once
and the eigenvalue $0$ occurs  $(n-1)^2$ times.
The entropy generated by the protocol can easily be calculated
from these eigenvalues.

\section{Entropy increase in a classical\\ one-way synchronization protocol}

\label{Bound}

Alice sends Bob a classical signal (``clock'') that is correlated with 
her clock. It is a classical physical system with non-trivial dynamics, i.e.,
a flow $\gamma_t:\Omega \rightarrow \Omega$ on its phase space $\Omega$.
Instead of sending such a clock Alice could also send Bob 
a composed system consisting  of the following two systems
which are easier to deal with:

\begin{enumerate} 
\item
A system which is described 
by the same
phase space $\Omega$  as the original system 
but with trivial 
time evolution, i.e, the measure $\mu$ is stationary in time and

\item 
a perfect classical clock which tells Bob exactly the time
that has been passed by since Alice has sent the message on the phase 
space $\Omega$. 
\end{enumerate}

Then Bob can implement the dynamical evolution
$\gamma_t$ corresponding to the original system.
For simplicity we assume the message space $\Omega$ to consist of
finitely many points  $\omega_1,\dots,\omega_l$. 
Since we shall derive a bound that is independent of
the message size we expect that it holds also in the limit
of infinite messages.

After Alice has sent the message we have a joint state on $\Omega$ and
$\cH_A$ such that Alice's clock is in the state $\rho_j$ if the message
contains the symbol $\omega_j$.
This case occurs with probability $p_j$. 
We  denote the joint state  by
\[
\sum_{j\leq l} p_j\, \rho_j \otimes \omega_j\,,
\]
keeping in mind that the left component of the tensor product
is a quantum density matrix and the right component a point 
in a classical space.
 
When Bob receives the message he may certainly perform an operation
on his quantum clock conditional on the message and throw
the message away.
However, this is not the most general operation. 
We allow also that he could, for instance,
implement a swap operation exchanging the state of the signal 
and his clock. To describe operations like this we have to change the 
point of view. From now on we consider the medium which carries the message
as a quantum system with Hilbert space $\cH_\Omega$ such that
$\omega_j$ are mutually orthogonal density matrices.
The orthogonality expresses the fact that only classical messages
are allowed.
At the time instant where the message arrives the joint state
of Alice and Bob is
\[
\nu:=\sum_j p_j \rho_j \otimes (\omega_j \otimes \eta)\,,
\]
where $\eta$ is an arbitrary density matrix of Bob's quantum clock.
Bob owns the two rightmost components and Alice the leftmost component
of the three-fold tensor product.
Regardless of the unitary operation that Bob performs on
$\cH'_B:=\cH_\Omega \otimes \cH_B$ the states $\omega_j \otimes \eta$ are
always transformed into mutually orthogonal states $\sigma_j$.
He may for instance apply some transformation $U$  
according to the
state of the classical clock. So far, we assumed that Alice 
and Bob implement only reversible 
transformations. Since we did not specify the physical systems
which they use this restriction does not imply any loss of generality.
After the protocol is finished the joint state 
$\nu$
evolves according to its autonomous dynamical evolution:
\[
(\alpha_t \otimes{\bf 1} \otimes \beta_t) (\nu)\,.
\]
We have emphasized that Alice and Bob are not allowed to keep 
the classical clock, they have to forget the time
in order to obtain a joint state of their quantum clocks
that is no longer correlated with the
 classical clock. A priori, it is not clear that the state
$\nu$ cannot be stationary and ``forgetting'' the time $t$ 
produces entropy. However, in the following we show
that $\nu$ cannot be stationary and prove a lower bound on
the entropy difference between $\nu$ and the time average
\[
\overline{\nu}:=\int_0^T 
(\alpha_t \otimes {\bf 1} \otimes \beta_t) (\nu) \, dt\,.
\] 
The idea is that every joint state that is prepared in a reversible
one-way protocol has a decomposition into product states which
are mutually orthogonal when they are restricted to Bob's 
system. On the other hand, there is no time-invariant joint state
with non-trivial synchronization with this  property.

In analogy to our results on 
the minimal entropy generation when timing information is read out from a
microscopic clock 
\cite{clockentropy}
our bound on the generated entropy relies on the energy bandwidth 
of the joint state. Explicitly the bandwidth is defined as follows.

\begin{Definition}
Let $(Q^A_r)_{r\in \R}$ and $(Q^B_r)_{r\in \R}$ 
the families of spectral 
projections of $H_A$ and $H_B$, respectively, i.e.,  
$Q^A_r$ projects onto the subspace
of $\cH_A$ that corresponds to eigenvalues not greater than $r$.

Let $\sigma$ be a joint state of Alice and Bob.
The bandwidth of Alice's clock 
is the least number $\Delta E_A$ such that
there exists $E \in \R$ such that
$(Q^A_{E+\Delta E_A} \otimes {\bf 1} -Q^A_E \otimes {\bf 1}) \sigma=\sigma$.
Define $\Delta E_B$ similarly.
\end{Definition}

Note that for finite spectral 
widths $\Delta E_A$ and $\Delta E_B$ 
the time evolution can equivalently be described by norm bounded Hamiltonians:
\[
H'_A:=Q_{E + \Delta E_A} Q_E H_A - {\bf 1} (E+ \Delta E_A/2)\,.
\]
Therefore we assume without loss of generality $\|H_A\| \leq \Delta E_A /2$
and $\|H_B\|\leq \Delta E_B/2$.

We will need the following Lemma:
\begin{Lemma}\label{Kull}
Let $W$ be a selfadjoint operator on a (not necessarily
finite dimensional) Hilbert space $\cW$ with discrete (not necessarily
finite) spectrum. Let $\rho$ be an arbitrary density matrix 
on $\cW$. Then the entropy difference between $\rho$ and
the average 
\[
\overline{\rho}:=\lim_{T\to\infty}\frac{1}{T}\int_0^T \exp(-iWt) \rho \exp(iWt)dt 
\]
is given by the Kullback-Leibler relative entropy
\[
K(\rho||\overline{\rho})=tr(\rho \ln \rho) -tr(\rho \ln \overline{\rho}) \,.
\]
\end{Lemma}

The proof is an immediate conclusion from
Lemma 1 in \cite{clockentropy} using the observation
that the average state $\overline{\rho}$ 
coincides with the post-measurement
state after measuring the observable $W$. Explicitly one has
\[
\overline{\rho}= \sum_j Q_j \rho Q_j \,,
\] 
where $Q_j$ are the spectral projections of $W$.
Note that discreteness of the spectrum ensures that the time average
exists \cite{Krengel}. Now we can state our main theorem:

\begin{Theorem}[Entropy generated by synchronization]${}$\\
\label{Haupt}
Let $\Delta t$ be the standard time deviation 
of the synchronism
$(\sigma,\alpha,\beta)$. Let $\Delta E$ be the total energy bandwidth
of $\sigma$, i.e., $\Delta E:=\Delta E_A + \Delta E_B$.
Then every classical one-way protocol 
to prepare the state $\sigma$ generates
at least the entropy 
\[
\Delta S= \frac{1}{16(\Delta E \Delta t)^2} \,.
\]
\end{Theorem}

\begin{Proof}
Let us modify our notation  for simplicity.
In contrast to
the definition above, we denote by $\beta_t$ the
joint time evolution of 
Bob's  clock and his memory (and not the dynamics of Bob's clock alone).
It acts on $\cH_B'$.
Given the time $t$ after the synchronization has been taken place
the joint state on $\cH_A\otimes\cH'_B$ is given 
by
\[
\nu=\sum_j p_j \, \rho_j \otimes \mu_j
\]
where $\mu_j$ are mutual orthogonal density matrices on $\cH_B$
due to the arguments above.

The resulting joint state $\sigma$ 
(which is the desired synchronism)
is obtained from $\nu$ by
forgetting the time $t$. It is hence
the time average $\sigma:=\overline{\nu}$.
First we consider
the trace norm distance between $\nu$ and $\overline{\nu}$: 
\begin{equation}\label{1normd}
\| \sum_j  p_j \rho_j \otimes \mu_j - 
\overline{\nu}
\|_1 \,.
\end{equation}
Let $A$ be an arbitrary observable with norm $1$.
With $H:=H_A \otimes {\bf 1} +{\bf 1}\otimes H_B$ we
have
 $\|[H,A]\|\leq 2 \|H\|=\Delta E$.

Note that
for any   two matrices $C,D$ one has $tr(C D) \leq \|C\|_1 \|D\|$
where $\|D\|$ denotes the usual operator norm of $D$.
Hence expression \ref{1normd} is at least
\[
\frac{tr (i[H,A] (\sum_j p_j \rho_j \otimes\mu_j 
- \overline{\nu}))}{\Delta E}
=\frac{tr(i[H,A] \sum_j p_j \rho_j \otimes\mu_j)}{\Delta E}\,.
\]
This equality is due to the time invariance of $\overline{\nu}$.

Now we choose observables $A_j$ with $\|A_j\|=1$ such that
\[
tr(-iA_j[H_A,\rho_j]=tr(i[H_A,A_j]\rho_j)= \|[H_A ,
\rho_j]\|_1\,.
\]
Let $P_j$ be a complete set of mutually orthogonal projections
separating the states $\mu_j$, i.e., $P_j\mu_j=\mu_j$ and $P_i\mu_j=0$ for 
$i\neq j$.
With the definition  $A:=\sum_j A_j \otimes P_j$ we obtain
\begin{eqnarray*}
tr(i[H,A] \sum_j p_j \rho_j \otimes\mu_j)&=&
tr(i[H,\sum_l A_l\otimes P_l] \sum_j p_j \rho_j \otimes\mu_j)\\
&=&\sum_{l,j}tr((i[H_A\otimes {\bf 1},A_l]\otimes P_l )(\rho_j \otimes \mu_j))
+\\&&\sum_{l,j} tr(i A_l\otimes [H_B,P_l] (\rho_j\otimes \mu_j)) \,.
\end{eqnarray*}

By simple calculations we find
that the only remaining term is
\[
\sum_j p_j\, tr(i[H_A,A_j]\rho_j)=
\sum_j p_j \|[H_A,\rho_j]\|_1\,.
\]
Using the bound 
\[
K(\nu||\overline{\nu})\geq \frac{\|\nu -\overline{\nu}\|^2 }{2}
\]
(see \cite{OhyaPetz})
 and Lemma \ref{Kull}
we conclude for the entropy generation
\begin{equation}\label{S1}
\Delta S \geq \frac{(\sum_j p_j \|[H_A,\rho_j]\|_1)^2}{2\Delta E}\,.
\end{equation}

On the other hand we know that the quality of synchronization can be defined
by 
\begin{equation}\label{Schluss}
\| \frac{d}{dt} \alpha_{t/2} \otimes \beta_{-t/2} (\sigma)\|_1=
\| [\frac{1}{2}(H_A\otimes {\bf 1} -{\bf 1}\otimes H_B), \overline{\nu}]\|_1\,.
\end{equation}
In the following we will use the abbreviation $\overline{C}$ 
for the ``time mean'' of an operator $C$ obtained by 
averaging over the evolution $\alpha_t \otimes \beta_t$.
The latter term in eq.~(\ref{Schluss})  
can be estimated as follows:
\begin{equation}\label{S2}
\| [\frac{1}{2}(H_A\otimes {\bf 1} -{\bf 1} \otimes 
H_B), \overline{\nu}]\|_1 = 
\| [H_A\otimes {\bf 1}, \overline{\nu}]\|_1 =
\|\overline{ [H_A\otimes {\bf 1}, \nu]}\|_1 \leq \| 
[H_A\otimes {\bf 1}, \nu]\|_1	 
\,.
\end{equation}
The last equality is due to the observation that 
the time evolution $\alpha_t \otimes\beta_t$ commutes with
the superoperator $[H_A\otimes {\bf 1},.]$. The inequality is due
to the fact that averaging over unitary dynamical evolution 
is a contractive map on the set of matrices.
We have
\[
\|[H_A\otimes {\bf 1}, \overline{\sum_j p_j \rho_j \otimes \mu_j}]\|_1 \leq
\| [H_A\otimes {\bf 1}, \sum_j p_j \rho_j \otimes \mu_j] \|_1
\leq \sum_j p_j \| [H_A,\rho_j]\|_1 \,.
\]
With inequality~(\ref{S1}) and (\ref{S2})  we conclude
\[
\Delta S \geq 
\frac{\|[\frac{1}{2}(H_A\otimes {\bf 1}-{\bf 1}\otimes H_B),
\overline{\nu}]\|_1^2}{4(\Delta E)^2}\,.
\]
With Lemma \ref{IntDelta} and eq.~(\ref{durchziehen}) 
we obtain
\[
\Delta S \geq \frac{1}{16(\Delta E \Delta t)^2}\,.
\]
\end{Proof}

So far, we used natural units, i.e., Planck's constant was 
assumed to be $1$. Using SI-units we obtain
\[
\Delta S \geq \frac{\hbar^2}{16 (\Delta E\Delta t)^2}\,
\]
as lower bound on the entropy generation
for a classical one-way protocol.

\section{Physical models for sending \\timing information by
classical communication}

\label{Explain}

Consider two microscopic clocks 
(like the systems in $\C^n$ considered in Section \ref{Obvious})
that should be synchronized. 
Both systems
are controlled by electronic devices
which are connected by an optical fiber. The fiber 
allows them to communicate
(see Fig.~1). 
We send a signal to one of 
both devices (or to both) which triggers the synchronization
procedure. Essential in our setting is that this signal 
is much less localized in time than $\Delta t$, the time accuracy 
of the synchronization.
Otherwise we deal
with a synchronization procedure that is run with {\it absolute} time.
Note that each signal with energy much less than  $\hbar/\Delta t$
satisfies this criterion by Heisenberg's uncertainty relation.
Say, for instance, the signal arrives at Alice's clock and 
starts the protocol. Then the device connected to Alice's clock
sends a light pulse to Bob carrying some information about the actual time
of clock A.

\begin{figure}
\centerline{
\epsfbox[0 0 309 153]{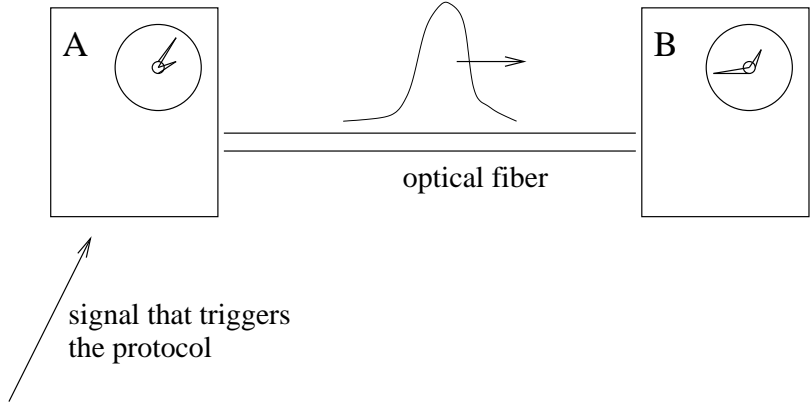}
}
\caption{Bob's device is triggered by the arrival of the signal and 
starts the clock $B$.}
\end{figure}

The optical fiber itself is a quantum channel.
Its quantum state may be described by a density matrix $\rho$ in
an appropriate Fock space with
time evolution
\[
\rho_t := \exp(-iH_L t) \rho \exp(iH_Lt)
\]
according to the corresponding Hamiltonian $H_L$ of the light field. 
The states
$\rho_t$ are necessarily non-commuting density matrices
since this holds for every non-trivial Hamiltonian
evolution
\cite{viva2002}. The timing information is hence necessarily 
to some extent quantum information.
However, now we include an assumption to the setting
which makes the protocol to be a classical communication protocol:
Assume that Bob's device starts his clock as soon as
the light signal arrives by measuring at every moment whether the light 
pulse has arrived or not. 

At first sight it seems that the entropy generation by this protocol
could simply be derived from results in \cite{clockentropy}. There we have
shown that every measurement that extracts timing information from a clock
with energy bandwidth $\Delta E$ produces at least the entropy 
\begin{equation}\label{absolute}
\Delta S\geq \frac{\hbar^2}{2(\Delta E \Delta t)^2}\,,
\end{equation}
whenever the measurement allows to determine the time up to an error
of $\Delta t$. Hence it seems that the measurement of the 
time of arrival measurement
must necessarily generate the entropy $\Delta S$ of inequality 
(\ref{absolute}). However, this  argument is not correct since we want to 
calculate the entropy increase from the point of view of somebody who
does not know the absolute time. From his point of view the quantum state
of the optical fiber is not the  state $\rho_t$ 
of the light field at a specific time instant $t$. It is rather a
mixture over all possible time instants. 

The following example shows that the entropy production 
in a classical synchronization protocol is not necessarily due to
a measurement but rather, as argued in Section \ref{Bound}, by the fact that
no absolute time is available.
Let Alice and Bob have  ``$n$-level'' clocks as in Section~\ref{Obvious}.
Let Alice's clock be in a maximally mixed state. Let Alice 
perform a measurement with respect to the Fourier basis
$|\psi_{t_j}\rangle$  with $t_j:=2\pi/n$ and
$|\psi_t\rangle := \exp(-iHt) |\psi\rangle$  with
$|\psi\rangle$ as in eq.~(\ref{Superp}).
This measurement does not generate any entropy since the clock
was already in its maximally mixed state before the measurement. 
If Alice writes the measurement
outcome into a memory that is sent to Bob we can continue with the protocol
as in Section~\ref{Obvious}. Then  entropy is  generated not earlier
than the moment where the correlation with the memory 
is lost in order to get a stationary joint state.

It may seem a little bit artifical to assume that
a classical clock is available {\it during} the 
transfer of timing information but has to be cleared after the protocol
has been finished.
In the situation of Fig.~1 the light signal may have large energy bandwidth
compared to the clocks of Alice and Bob. Hence it may be considered
approximatively as a classical clock. However, this signal is absorbed
by Bob's device. This does not necessarily mean that the signal energy itself
is lost. The device may be designed in an energy-saving way such that
the signal energy is used to reload a capacitor
(by a solar cell, for instance). The fact that the signal
is absorbed by Bob's device means indeed that the classical 
clock used for the time transfer is {\it no longer available}.
Of course it is not necessary that the light signal is absorbed.
It could also be captured by a cavity such that the cavity 
contains an oscillating light field. But in this case 
Bob's clock should be considered as the 
composed system 
consisting of the cavity and Bob's original clock.
Anyway, the classical clock which is approximatively given by a 
classical light field in the optical fiber, is no longer existent after 
the light pulse has arrived. 
The authors think that this shows that the setting
presented here is rather natural.

\section{Cost for resetting synchronized clocks:\\ 
 quantum discord}

\label{Discord}

As already mentioned in Section \ref{Adv} quantum communication may also
be advantageous when the synchronized clocks should be reset.
This is shown in the following example. Let Alice and Bob each have a 
``two-level clock'', i.e., two-level systems in the state
\[
|\psi_t\rangle:=\frac{1}{\sqrt{2}}(|0\rangle +\exp(-i\omega t)|1\rangle )\,.
\]
Such a clock is reset, for instance, when it is set into the
state $|0\rangle$ (one may think of a stop-watch which was running for 
a while and should be stopped and reset afterwards).
The problem is that neither Alice nor Bob is able to perform
a thermodynamically reversible operation which converts
$|\psi_t\rangle$ into $|0\rangle$ when no additional clock
is available since they do not know $t$.
However, Alice and Bob could each have an ancilla qubit which is
a degenerated two-state system with the zero operator as Hamiltonian. 
Let these qubits be initialized to $|0\rangle$.
Then both of them
may exchange the state of the two-level system with the 
state of the degenerated system. 
Obviously, both clocks are set to $|0\rangle$ by this 
procedure. However, in case Alice and Bob have not agreed upon a 
common  time instant where they perform the operation this 
whole process is irreversible since the correlation between
Alice's and Bob's clock is lost.
From the point of view of the ignorant observer the mixture 
of $|\psi_t\rangle \otimes |\psi_t\rangle$ over all $t$
is transformed into an uncorrelated state where Alice's and Bob's
ancilla qubits are both in 
the
maximally mixed state.
But, in analogy to the requirements
listed in Section~\ref{Obvious}, we allow Alice and Bob to transfer a
 classical 
clock. However,  
we assume that the synchronized quantum clocks are not correlated
with the classical clock when the resetting procedure
is {\it started}. Using classical clock transfer they can synchronize
their ``clock resetting'' process and assure that the joint state of their
ancilla qubits is a mixture of all states $|\psi_t\rangle \otimes
|\psi_t\rangle$. As long as they keep this correlated joint state 
they have indeed reset their clocks in a reversible way.
However, this is an unsatisfactory end of the process:
Maybe Alice would
like to synchronize her clock afterwards with a third party, say 
Carol and would like to reset this synchronization  after a while
in order to  synchronize with Dave.
Then she would need  an additional ancilla qubit for each party in order 
to keep the correlations. To avoid those unrestricted resource requirements
we would like to resolve all correlations between Alice's and Bob's
clocks or ancillas. When quantum communication is allowed Bob may for instance
have two initialized ancilla qubits and the mixture of all states
$|\psi_t\rangle \otimes |\psi_t\rangle$ may be converted 
into the same mixture in Bob's two ancilla qubits. 
But here we will only allow  classical one-way communication and
show that resolving the correlations between ancillas or clocks
will unavoidably lead to dissipation. We will show this by proving
that the joint state of synchronized microscopic clocks 
has always quantum discord (``quantum correlations without 
entanglement'' \cite{ZurekDiscord}). 
Let us explain briefly this concept introduced by Ollivier and Zurek.
For two classical systems $A$ and $B$ (formally described by random variables
 $A$ and $B$) one may define mutual information  
in two equivalent ways
\cite{Cover}:
\begin{enumerate}

\item
The symmetric expression 
\[
I(A:B):=H(A) +H(B) -H(A,B)\,,
\]
where $H(A)$, $H(B)$, $H(A,B)$ 
are the Shannon entropies of $A$ , $B$ , or joint entropy of $A$ and 
$B$, respectively.

\item
The asymmetric expression
\[
I(A:B):= H(B) -H(B|A)\,,
\] 
where $H(B|A)$ is the entropy of $B$ given $A$.

\end{enumerate}

The quantum analogue of 1. is given by
\begin{equation}\label{Mutual}
S(\sigma^A)+S(\sigma^B)-S(\sigma)
\end{equation}
 where $S(.)$ denotes the von-Neumann entropy
\cite{OhyaPetz}, $\sigma$ is the joint density matrix of the system $A$ and $B$
and $\sigma^A$ and $\sigma^B$ are the restrictions of the state 
$\sigma$ to
the subsystem $A$ and $B$, respectively.

The analogue of 2. refers to measurements on the system $A$. 
As in \cite{ZurekDiscord} 
we restrict our attention to von-Neumann measurements
described by a family $(P_j)$ of orthogonal projections acting on
Alice's Hilbert space \footnote{In our setting this is no loss of generality
since one may count the quantum clock together with
arbitrarily many ancillas as a new clock.}.
Define the probabilities 
\[
p_j := tr((P_j \otimes 1) \sigma)
\]
and the selected post-measurement states
\[
\sigma_j := (P_j \otimes 1) \sigma (P_j \otimes 1)/p_j\,.
\]
Then the entropy of $B$ given the measurement outcome is given by
\[
\sum_j p_j S(\sigma_j^B)
\]
and the difference 
\[
S(\sigma^B) -\sum_j p_j S(\sigma_j^B)
\]
may be considered as the quantum analogue of 2.

The {\it discord} $\partial(B|A)$ 
 as introduced by Ollivier and Zurek
 \cite{ZurekDiscord}
is the minimum of  all values 
$\partial_{(P_j)}(B|A)$ over all measurements $(P_j)$, with
\[
\partial_{(P_j)} (B|A):= S(\sigma^A)-S(\sigma)+\sum_j p_j S(\sigma_j^B))\,.
\]
This quantity is the difference between both possible translations
of mutual entropy.

A rather artificial combination of 1. and 2. leads to
\begin{equation}\label{third}
I(A:B)=H(A) +H(B) -(H(A) +H(B|A))\,.
\end{equation}
Note $H(A)$ has two possible translations
into the quantum setting. It may either be the entropy of $\rho^A$ 
or of the unselected post-meassurement $\sum_j p_j \sigma_j ^A$.
As Zurek noted \cite{ZurekDemon}
one may also choose the first possibility for the first term $H(A)$ 
and the second  possibility for the second term $H(A)$.
This leads to 
\begin{equation}\label{thirdterm}
S(\sigma^A)+S(\sigma^B) -S(\sum_jp_j\sigma_j^A) -\sum_j 
p_jS(\sigma_j ^B)\,.
\end{equation}
In \cite{ZurekDemon} Zurek considered the difference between
expression~(\ref{Mutual}) 
 and expression (\ref{thirdterm}): 
\begin{equation}\label{Dis2}
\delta_{(P_j)}(B|A):=S(\sum_j p_j \sigma_j^A) +\sum_j p_j S(\sigma_j^B)
-S(\sigma)\,.
\end{equation}
He called the minimum of all values $\delta_{(P_j)}(B|A)$ over all
measurements
also {\it discord} and denoted it by the symbol
$\delta(B|A)$.
He showed this expression of discord to be thermodynamically
relevant \cite{ZurekDemon}. It is the difference between the 
entropy cost for erasing the joint state of a bipartite 
quantum memory when  only classical one-way communication from
Alice to Bob is allowed to the erasure cost in
optimal quantum protocols.
In our setting it is the difference between the 
entropy that has to be transferred to the environment
when Alice and Bob reset their synchronized clocks
using classical communication to the amount they would have to transfer
to the environment if quantum communication was allowed. 
We will prove a lower bound on the discord of two synchronized 
quantum clocks.
The following Lemma will be useful in our proof:

\begin{Lemma}\label{DistDis}
The expression $\delta_{(P_j)}(B|A)$ as in
eq.~(\ref{Dis2}) 
 associated with a 
measurement $(P_j)$ can be written as a sum of Kullback-Leibler
distances:
\begin{eqnarray*}
\delta_{(P_j)}(B|A)&=&K(\sigma || \sum_j p_j \sigma_j) +
\sum_j p_j K(\sigma_j || \sigma_j ^A \otimes \sigma_j^B)\\&=&
K(\sigma || \sum_j p_j \sigma_j) +
K(\sum_j p_j\sigma_j || \sum_l p_l\sigma_l ^A \otimes \sigma_l^B)
\,.
\end{eqnarray*}
\end{Lemma}

This expression has a rather intuitive meaning:
The first summand is the distance between the pre-measurement state
and the unselected post-measurement state (as mentioned in the remarks
after Lemma \ref{Kull} 
this coincides with the entropy generated by the measurement). 
The second term is the average distance between the selected 
joint state and the tensor product of the reduced (selected) post-measurement
states. If a joint state has discord this means that each measurement either
generates entropy or it does not resolve the correlations, i.e.,
the selected post-measurement state is still correlated. 
This suggests already that correlations with discord cannot be
resolved in a thermodynamically reversible way by measurements of one party.

\vspace{0.3 cm}
\begin{Proof} (of Lemma \ref{DistDis}):
Obviously one has
\begin{eqnarray*}
\delta_{(P_j)}(B|A)&=& S(\sum_j p_j \sigma_j^A) +\sum_j p_j S(\sigma_j ^B)
-S(\sum_j p_j \sigma_j) +S(\sum_j p_j \sigma_j) -
S(\sigma)\\&=&
S(\sum_j p_j \sigma_j^A) +\sum_j p_j S(\sigma_j ^B)
-S(\sum_j p_j \sigma_j) + K(\sigma || \sum_j p_j \sigma_j)
\,.
\end{eqnarray*}
Since the states $\sigma_j$ are orthogonal and also
their restrictions to $A$ are orthogonal we have
\[
S(\sum_j p_j \sigma_j)= \sum_j p_j S(\sigma_j) +H(p)
\]
and
\[
S(\sum_j p_j \sigma_j^A)= \sum_j p_j S(\sigma_j^A) +H(p)\,,
\]
where $H(p)$ is the amount of information of the measurement result.
We conclude
\[
\delta_{(P_j)}(B|A)=
\sum_j p_j S(\sigma_j^A) +\sum_j p_j S(\sigma_j ^B)
-\sum_j p_j S(\sigma_j) + K(\sigma || \sum_j p_j \sigma_j)\,.
\]
Using the identity
\[
tr(\sigma_j \ln (\sigma_j^A \otimes \sigma_j^B))=
tr((\sigma_j^A \otimes \sigma_j^B) 
\ln (\sigma_j^A \otimes \sigma_j^B)
\]
we obtain
\[
\delta_{(P_j)}(B|A)=\sum_j p_j K(\sigma_j || \sigma_j^A \otimes \sigma_j ^B)
+ K(\sigma || \sum_j p_j \sigma_j)
\]
by elementary calculation.

The last equality of the statement, namely
\[
\sum_j p_j K( \sigma_j || \sigma_j ^A \otimes \sigma_j^B)
=  K( \sum_j p_j \sigma_j ||
\sum_j p_j  \sigma_j ^A \otimes \sigma_j^B)\,,
\]
is easy to check
since the states $\sigma_j$ and $\sigma_j^A \otimes \sigma_j ^B$
act on the images of the projections $P_j\otimes 1$, i.e.,
they act on mutually orthogonal subspaces for different $j$. 
\end{Proof}

Now we shall prove a lower bound on the quantum discord
of the joint state of two synchronized clocks.

\begin{Theorem}[Discord of synchronized clocks]${}$\\
Let $(\rho,\alpha,\beta)$ be a synchronism of quantum clocks
with equal period.
Then the quantum discord between $A$ and $B$ is non-vanishing 
and we
have
\[
\delta( A|B) \geq \frac{1}{256(\Delta t \Delta E)^2}\,\,\,\,\hbox{ and } \,\,\,\,\,
\delta (B|A) \geq \frac{1}{256(\Delta t \Delta E)^2}\,,
\]
where $\Delta t$ is the 
accuracy of the synchronization, i.e., the
standard time deviation and $\Delta E$ is the energy bandwidth of the clocks. 
\end{Theorem}

\begin{Proof}
Due to the symmetry with respect of $A$ and $B$ it is sufficient to prove 
the second inequality.
We prove the bound by  showing that it holds for every von-Neumann 
measurement
$(P_j)$.
We define
\[
d_1:= \|\sigma - \sum_j p_j \sigma_j\|_1
\]
and
\[
d_2:= \|\sum_jp_j \sigma_j  -\sum_j \sigma_j^A \otimes 
\sigma_j^B \|_1\,.
\]
The idea of the proof is that  $d_1$ and $d_2$ 
cannot be simultaneously small. Otherwise the joint state $\sigma$ 
of the synchronized clocks would be
close to the state
\[
\sum p_j \sigma_j^A \otimes \sigma_j^B\,.
\]
This state consists of  product states that
are locally distinguishable on Alice's subsystem.
The fact that no time invariant state with
non-trivial synchronization can have this property was already the key
idea in the proof of Theorem \ref{Haupt}.

Now we have
\begin{eqnarray*}
\|[{\bf 1} \otimes H_B,\sigma]\|_1 
&\leq & \|[{\bf 1} \otimes 
H_B, \sum_j p_j \sigma_j ^A \otimes \sigma_j^B ]\|_1 +(d_1 +d_2)
\Delta E \\ &=&
\sum_j p_j \|[{\bf 1} \otimes H_B,  \sigma_j ^A \otimes \sigma_j ^B ]\|_1 +(d_1 +d_2)
\Delta E\,.
\end{eqnarray*}
The first inequality follows from the fact that $H_B$ is bounded
with $\|H_B\|\leq \Delta E/2$.
Therefore the commutator with $\sigma$ cannot differ from the commutator
with 
\[
\sum_j p_j \sigma_j ^A \otimes \sigma_j^B
\]
by more  than the amount $(d_1+d_2)\Delta E$.
The last equality is due to the orthogonality of the states
$\sigma_j^A$ since ${\bf 1}\otimes H_B$ acts only on
the second tensor component.

In analogy to the proof of Theorem \ref{Haupt} we choose 
observables $A_j$ such that 
\[
tr(i \,[H_B,A_j] \sigma_j ^B)  =  
 \|[H_B, \sigma_j^B]\|_1 \,,
\]
and define $A:=\sum_j P_j \otimes A_j$, where $P_j$ are the measurement 
operators.

Some calculations show 
\begin{eqnarray*}
\|[{\bf 1}\otimes H_B,\sum_j p_j \sigma_j^A\otimes \sigma_j^B]\|_1
&=& tr(i[{\bf 1}\otimes H_B, A] \sum_j p_j \sigma_j^A\otimes \sigma_j^B)\\
&=& tr(i[H,A] \sum_j p_j \sigma_j^A\otimes \sigma_j^B)\,.
\end{eqnarray*}

Since the difference between  
\[
tr(i[H,A]\sum_j p_j \sigma_j^A\otimes \sigma_j^B]
\]
and 
\[
tr(i[H,A]\sigma)
\]
cannot be greater than $(d_1+d_2)\Delta E$ according to the same
arguments as above we have
\begin{eqnarray*}
\|[{\bf 1}\otimes H_B,\sigma]\|_1 &\leq& 
 tr(i\,[H, A] \sigma)  +(d_1+d_2)2\Delta E \\
&=& 
(d_1+d_2)\,2\Delta E \,.
\end{eqnarray*}
The last equality is due to the stationarity of the state $\sigma$.
Due to the inequality  
\[
K(\gamma||\tilde{\gamma}) \geq \frac{\| \gamma-\tilde{\gamma}\|^2}{2}
\]
we have 
\[
\delta_{(P_j)}(B|A) \geq \frac{d_1^2 +d_2^2}{2} 
\]
With 
\[
d_1+d_2 \geq \frac{|[{\bf 1}\otimes H_B,\sigma]\|_1}{2\Delta E}
\]
we have
\[
d_1^2 +d_2^2 \geq \frac{|[{\bf 1}\otimes H_B,\sigma]\|^2_1}{8(\Delta E)^2}\,.
\]
We conclude
\[
\delta_{(P_j)}(B|A)
\geq \frac{\|[{\bf 1}\otimes H_B,\sigma]\|^2_1}{16(\Delta E)^2}\,.
\]
Using Lemma \ref{IntDelta} and eq.~(\ref{durchziehen})
we conclude 
\[
\delta(B|A) \geq \frac{1}{256(\Delta E \Delta t)^2}\,.
\]
\end{Proof}

\section{Implications for low power computation}

\label{Imply}

When we discuss hypothetical low power computers here we mean
devices with energy consumption much below the consumption
of any present technology or prototypes for the middle future.
Nevertheless we find it worth to discuss under which circumstances
fundamental lower bounds on the power consumption of computers
can be proved. Currently, the only fundamental bound that is known 
is Landauer's principle \cite{Landau} stating that every logical
irreversible computation leads unavoidably to power consumption.
The converse statement that power consumption could in principle 
be avoided at all by
using logically reversible circuits, is questionable.
It is well-known that all classical computations can be implemented
using Toffoli-gates \cite{Toffoli} which can be considered as unitary 
transformations in a Hilbert space of the computer due to their
logical reversibility.  
However, the signals controlling
the
implementation time is always excluded in the thermodynamical
considerations \cite{viva2002}. This is correct if the 
signal energy is sufficiently high such that the signal 
can  be considered  classical.
If its quantum nature is taken into account severe problems
with thermodynamical reversibility may appear (some thoughts on this
problem can be found in \cite{viva2002}).
To our knowledge, the only theoretical models for
a  closed physical system can be found in \cite{Benioff,Feyn86,Marg86,Marg}.
The model in \cite{Benioff} uses a  Hamiltonian which is unbounded
below. Such Hamiltonians  only exists in the limit  
of high system energy since the energy is then much above 
the ground state. The concept of \cite{Marg86,Marg}  
avoids a global clocking
mechanism at all. It leads to states which are superpositions
of different results. The register is highly entangled, i.e.,
a lot of quantum information is transferred among different parts
of the register.

Now we show in which way our results may give lower bounds
for the energy consumption of all computers that rely on too
{\it conventional} concepts, for instance, in the sense
that they do not transfer quantum coherent signals. 
Since our results may apply to different levels
of a computer (communication between 
transistors, devices, gates, processors, computers) 
we will not specify the
components at all.

Consider two components $A$ and $B$  
each producing an output $a_j$ and $b_j$ 
in time step $j$.  A third logical device $C$ receives 
$a_j$ and $b_j$ as inputs (see Fig.~2). 

\begin{figure}
\centerline{
\epsfbox[0 0  251 207]{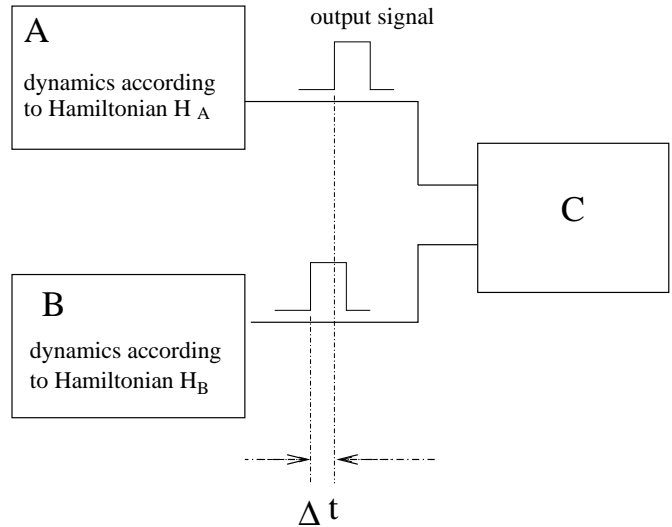}
}
\caption{Two components $A$ and $B$ 
that have to be synchronized. When their evolution 
can  approximatively be described 
by separate Hamiltonians our lower bounds
on the required synchronization entropy are valid.}
\end{figure}

Hence the components $A$ and $B$ have 
to be synchronized up
to an accuracy $\Delta t$, the length of the time steps. 
Assume $A$ and $B$  to be quantum systems that evolve approximatively
according to their Hamiltonians $H_A$ and $H_B$.
Of course, this can only be a rough approximation since 
both systems receive signals from other components and send
signals to $C$. Nevertheless we tend to  
believe that our results above suggests that the required synchronization 
requires  either quantum communication between $A$ and $B$ or leads 
unavoidably to power consumption. 
Although we were only able to prove our bounds for one-way protocols
it seems likely that also classical  multi-step protocols
generate some entropy.

We admit that arguments like this should  be analyzed thoroughly.
This should be subject of further research.

Thanks to Khoder El-Zein for some useful calculations
and Thomas Decker for helpful remarks.
This work has been supported by grants of the DFG-SPP VIVA
project No. Be 887/12.


\begin{thebibliography}{10}

\bibitem{BennettNonlocal}
C.~Bennett, D.~DiVincenzo, C. Fuchs, T.~Mor, E.~Rains, P.~Shor, J.~Smolin, and
  W.~Wootters.
\newblock Quantum nonlocality without entanglement.
\newblock {\em LANL-preprint quant-ph/9804053}.

\bibitem{ZurekDiscord}
H.~Ollivier and W.~Zurek.
\newblock Quantum discord: A measure for the quantumness of correlations.
\newblock {\em Phys. Rev. Lett.}, 88:017901, 2002.

\bibitem{ZurekDemon}
W.~Zurek.
\newblock Quantum discord and Maxwell's demons.
\newblock {\em Phys. Rev. A}, 67:012320, 2003.

\bibitem{clock}
D.~Janzing and T.~Beth.
\newblock Quasi-order of clocks and their synchronism and quantum bounds for
  copying timing information.
\newblock {\em IEEE Trans. Inform. Theor}, 49(1):230--240, 2003.

\bibitem{Gruenbaum}
B.~Gr\"{u}nbaum.
\newblock {\em Convex Polytopes}.
\newblock John Wileys \& Sons, London, 1967.

\bibitem{clockentropy}
D.~Janzing and T.~Beth.
\newblock Bounds on the entropy generated when timing information is extracted
  from microscopic systems.
\newblock {\em LANL-preprint quant-ph/0301125}.

\bibitem{Krengel}
U.~Krengel.
\newblock {\em Ergodic Theory}.
\newblock Walter de Gruyter, Berlin, 1985.

\bibitem{OhyaPetz}
M.~Ohya and D.~Petz.
\newblock {\em Quantum entropy and its use}.
\newblock Springer Verlag, 1993.

\bibitem{viva2002}
D.~Janzing and Th. Beth.
\newblock Are there quantum bounds on the recyclability of clock signals in low
  power computers?
\newblock In {\em Proceedings of the DFG-Kolloquium VIVA}, Chemnitz, 2002.
\newblock {\em LANL-preprint quant-ph/0202059}.

\bibitem{Cover}
T.~Cover and J.~Thomas.
\newblock {\em Elements of Information Theory}.
\newblock Wileys Series in Telecommunications, New York, 1991.

\bibitem{Landau}
R.~Landauer.
\newblock Irreversibility and heat generation in the computing process.
\newblock {\em IBM Res. J.}, pages 183--191, July 1961.

\bibitem{Toffoli}
T.~Toffoli.
\newblock Reversible computing.
\newblock {\em MIT Report MIT/LCS/TM-151}, 1980.

\bibitem{Benioff}
P.~Benioff.
\newblock The computer as a physical system: A microscopic quantum mechanical
  model of computers as represented by turing machines.
\newblock {\em J. Stat. Phys.}, 22(5), 1980.

\bibitem{Feyn86}
R.~Feynman.
\newblock Quantum mechanical computers.
\newblock {\em Found. Phys.}, 16(6):503--531, 1986.

\bibitem{Marg86}
N.~Margolus.
\newblock Quantum computation.
\newblock {\em Ann. NY. Acad. Sci.}, 480(480-497), 1986.

\bibitem{Marg}
N.~Margolus.
\newblock Parallel quantum computation.
\newblock In W.~Zurek, editor, {\em Complexity, entropy, and the physics of
  information}, volume VIII, pages 273--287. Santa Fee Institute, Adison
  Wesley, 1990.
\newblock SFI-studies.

\end{thebibliography}
\end{document}